\documentclass[aps,prx,reprint,superscriptaddress,amsmath,amssymb,nofootinbib]{revtex4-2}
\usepackage[T1]{fontenc}
\usepackage{lmodern}
\usepackage[british]{babel}
\usepackage{amsthm,mathtools}
\usepackage{microtype}
\usepackage[colorlinks=true,linkcolor=blue,citecolor=blue,urlcolor=blue]{hyperref}

\newtheorem{theorem}{Theorem}
\newtheorem{prop}{Proposition}
\DeclareMathOperator{\Tr}{Tr}
\DeclareMathOperator{\Ch}{Ch}
\DeclareMathOperator{\Ad}{Ad}
\newcommand{\id}{\mathrm{id}}

\begin{document}
\title{Inevitability of the Petz transpose map}
\author{Jacopo Surace}
\email[Corresponding author: ]{jacopo.surace@gmail.com}
\affiliation{Aix-Marseille University, CNRS, LIS, Marseille, France}

\author{Chenxu Li}
\affiliation{Center for Quantum Information, Institute for Interdisciplinary Information Sciences (IIIS), Tsinghua University, Beijing 100084, China}

\author{Francesco Buscemi}
\email[Corresponding author: ]{buscemi@nagoya-u.jp}
\affiliation{Dept. Mathematical Informatics, Nagoya University, Nagoya 464-8601, Japan}

\begin{abstract}
Bayes' rule is central to classical statistical inference: given a prior and a stochastic model, it determines the corresponding retrodictive conditional probabilities. In quantum theory, however, the question of whether an equally general rule of retrodiction exists and, if so, whether it is uniquely determined remains nontrivial. Here we prove that a retrodiction assignment defined for all priors and channels, which takes each predicted output state back to its prior, must coincide with the Petz transpose map if it is involutive, respects sequential and parallel composition, and agrees with unitary inversion. The theorem holds for channels between finite-dimensional full matrix algebras, with the input and output spaces restricted to the supports of the corresponding reference states. In particular, affinity under probabilistic mixtures and the characteristic square-root dependence on the reference states follow from these requirements rather than being postulated. Without involutivity, the remaining requirements characterize a family of doubly rotated Petz maps. Our result thus explains why the Petz transpose map arises in seemingly different areas of quantum theory: it is the unique quantum analogue of Bayes' rule compatible with these consistency requirements.
\end{abstract}
\maketitle

\emph{Introduction.}
Bayes' rule allows a statistical model to be used in both directions: from a hypothesis to its predicted consequences, and from an observed consequence to what can be inferred about the hypothesis. When the model describes a physical process, this change in the direction of inference does not require the process itself to be reversible. Indeed, information lost during the evolution cannot in general be recovered; nevertheless, a prior and the forward transition probabilities determine a reverse conditional probability. The question about the ``inevitability'' of Bayes' rule has motivated derivations from consistency requirements on probabilistic reasoning and from principles governing the incorporation of new information~\cite{DeFinetti1937,Cox1946,Williams1980,DiaconisZabell1982,Jaynes2003}. These approaches concern different assumptions, but share the aim of identifying the conditions under which Bayesian inference is necessary.

The corresponding question in quantum theory involves a further difficulty. A classical prior and a stochastic process determine a joint probability distribution, from which either direction of inference can be obtained by conditioning. A quantum state and a channel, on the other hand, do not come with an equally unambiguous joint-state interpretation. Quantum conditional states connect Bayesian inference with channel--state duality~\cite{Leifer2006,LeiferSpekkens2013,ParzygnatFullwood2023}, while axiomatic studies of states over time identify limitations and constructions under different requirements~\cite{HorsmanEtAl2017,FullwoodParzygnat2022,LieNg2024,ParzygnatFullwoodBuscemiChiribella2024}. Alternatively, one may seek to determine the reverse process through a variational principle. In the quantum minimum-change principle, for instance, a reverse channel is selected by maximizing the fidelity between bipartite states representing the forward and reverse processes~\cite{BaiBuscemiScarani2025}. However, this formulation still requires a choice of process representation, together with a figure of merit that specifies what is meant by minimum change. The question then arises whether quantum retrodiction can instead be determined directly by consistency requirements on the reversal of channels.

Our starting point is the axiomatic framework of Ref.~\cite{ParzygnatBuscemi2023}, which expresses precisely such requirements: retrodiction should respect sequential and parallel composition, return the original process when applied twice, and agree with ordinary inversion for unitary dynamics. The assignment defined by the Petz transpose map~\cite{Petz1986,Petz1988} satisfies these requirements, but whether they uniquely determine it was left open. The corresponding uniqueness question also appears in the recent extension to von Neumann algebras~\cite{KarmakarParzygnat2026}.

Here we prove that these four requirements uniquely determine the Petz assignment for channels between finite-dimensional full matrix algebras. For each prior and forward channel, retrodiction must take the predicted output state back to that prior, without being required to undo the dynamics on arbitrary input states. For a prior $\pi$ and a channel $\Phi$, we call $\pi$ and its predicted output $\sigma=\Phi(\pi)$ the input and output reference states, respectively, and $\pi\to\sigma$ the reference transition. A single retrodiction assignment must provide these reversals for every prior and channel, while satisfying the four requirements consistently across reference transitions and systems. We also show that, when involutivity is removed, the remaining three requirements characterise a family of doubly rotated Petz maps, with rotations chosen consistently across reference states and systems. We give the operational argument and explain the mechanism of uniqueness below; the complete proof is in the Supplemental Material.

\ 

\emph{Consistency of retrodiction.}
For a classical transition probability $p(y|x)$ and a prior $\pi(x)$, assuming strictly positive marginals, Bayes' rule assigns the reverse conditional probability
\begin{equation}
\widehat p(x|y)=\frac{\pi(x)p(y|x)}{\sigma(y)},
\label{eq:classical-bayes}
\end{equation}
where $\sigma(y)=\sum_x\pi(x)p(y|x)$. Averaging the reverse conditional probabilities with the predicted distribution gives $\sum_y\widehat p(x|y)\sigma(y)=\pi(x)$. Thus Bayes' rule reverses the reference transition from the prior to its prediction, even when the forward process is not invertible. Reversing twice returns the original transition.

To formulate the corresponding requirements in quantum theory, let $\mathcal{B}(\mathcal{H})$ denote the algebra of all operators on a finite-dimensional Hilbert space $\mathcal{H}$. Let $\Ch(\pi,\sigma)$ denote the set of completely positive trace-preserving (CPTP) maps, or \emph{channels}, $\Phi:\mathcal{B}(\mathcal{H}_A)\to\mathcal{B}(\mathcal{H}_B)$ such that $\Phi(\pi)=\sigma$, where $\pi$ and $\sigma$ are full-rank states. A retrodiction assignment is a family of maps
\begin{equation}
\mathcal R_{\pi\to\sigma}:\Ch(\pi,\sigma)\longrightarrow\Ch(\sigma,\pi)
\end{equation}
defined for every full-rank reference pair on arbitrary finite-dimensional systems, including ancillary systems and the one-dimensional system.
This formulation is equivalent to specifying a channel $\Phi$ and a prior $\pi$, since the predicted state is then fixed by $\sigma=\Phi(\pi)$. The sets $\Ch(\pi,\sigma)$ group channels with the same prior and prediction, and considering all reference pairs introduces no additional restriction.
 The target $\Ch(\sigma,\pi)$ enforces $\mathcal R_{\pi\to\sigma}[\Phi](\sigma)=\pi$, so the prescribed reference transition is reversed. This does not require the reverse channel to undo $\Phi$ on other states. The assigned reverse channel may depend on the prior even when the forward channel is unchanged. When the reference states are understood, we write simply $\mathcal R[\Phi]$.

 We impose the following four conditions on this entire family, over all reference pairs and systems.

\emph{(i) Involutivity.}
Reversing the direction of inference twice must return the original process, $\mathcal R[\mathcal R[\Phi]]=\Phi$. 

\emph{(ii) Sequential composition.}
If $\Psi$ follows $\Phi$, retrodiction reverses their order: $\mathcal R[\Psi\circ\Phi]=\mathcal R[\Phi]\circ\mathcal R[\Psi]$. 
The reverse of a composite channel can therefore be obtained by composing the individual reverse channels in the opposite order, provided their reference transitions match.

\emph{(iii) Parallel composition.}
For two processes acting independently, with product reference states, $\mathcal R[\Phi\otimes\Psi]=\mathcal R[\Phi]\otimes\mathcal R[\Psi]$. Adding an independent system should not affect how inference is reversed on the original one.

\emph{(iv) Unitary inversion.}
For a unitary channel $\Ad_U(X)=UXU^\dagger$, we require $\mathcal R[\Ad_U]=\Ad_{U^\dagger}$, for every full-rank prior. Since unitary evolution loses no information, its reverse is the ordinary inverse, independently of the reference state chosen.

Notice that involutivity has a different status from the other three requirements. While sequential and parallel composition, together with unitary inversion, express compatibility with the operational structure of quantum circuits, involutivity imposes an additional symmetry between the two directions of inference. As we shall see, it is precisely this symmetry that eliminates the remaining freedom in the retrodiction assignment (see Proposition~\ref{lem:without-involutivity-main}).

\begin{theorem}[Uniqueness of Petz retrodiction]\label{thm:uniqueness-main}
There is exactly one retrodiction assignment on arbitrary finite-dimensional systems with full-rank reference states satisfying (i)--(iv), namely the Petz assignment,
\begin{equation}
\mathcal R_{\pi\to\sigma}[\Phi](X)=\sqrt{\pi}\,\Phi^\dagger\!\left(\sigma^{-1/2}X\sigma^{-1/2}\right)\sqrt{\pi},
\label{eq:petz-main}
\end{equation}
where $\Phi^\dagger$ is the trace adjoint.
\end{theorem}

The theorem also covers arbitrary priors and predictions, including pure states, when the input and output spaces are restricted to their respective supports.
Let $P$ and $Q$ be the orthogonal projections onto the supports of $\pi$ and $\sigma=\Phi(\pi)$, respectively.
 Every positive $X=PXP$ satisfies $X\leq c\pi$ for some $c\geq0$, so positivity gives $0\leq\Phi(X)\leq c\sigma$. Hence $\Phi$ restricts to a channel from $\mathcal B(P\mathcal H_A)$ to $\mathcal B(Q\mathcal H_B)$, on which both references are full-rank. Restriction respects sequential and parallel composition and unitary inversion, so the theorem applies on these reduced systems. 
Accordingly, for arbitrary priors we understand channels as maps between the reference supports and impose involutivity on those maps. As in classical Bayesian inference, no reverse prescription is thereby fixed for observations excluded by the predicted state. In particular, if the input reference is full-rank, restricting the output space preserves the entire forward channel even when its predicted state is singular. The Supplemental Material also relates this support convention to the formulation of retrodiction for states that are not full-rank in Ref.~\cite{Parzygnat2024}.

\ 

\emph{Retrodiction of probabilistic mixtures.}
We first show that a random choice of channels sharing a reference transition is reversed with the same probabilities. The starting observation is that retrodiction exchanges preparation and discarding. Indeed, $\mathcal S_\rho(z)=z\rho$ and $\Tr$ are the only channels in $\Ch(1,\rho)$ and $\Ch(\rho,1)$, respectively, hence
\begin{equation}
\mathcal R[\mathcal S_\rho]=\Tr,\qquad \mathcal R[\Tr]=\mathcal S_\rho.
\label{eq:preparation-trace-main}
\end{equation}

Consider now $\Phi_0,\Phi_1\in\Ch(\pi,\sigma)$, selected with probabilities $1-t$ and $t$, where $0<t<1$. Unitary swaps controlled by an independent random bit route the input through the selected channel and an auxiliary system prepared in $\pi$ through the other. Both channels act in parallel, after which the selected output is retained and the remaining systems are discarded. Since the channels share a reference transition, the swaps preserve the prescribed reference states, so the composition and unitary-inversion requirements allow us to reverse the implementation operation by operation. Preparation and discarding are exchanged, so the control is prepared with its original probabilities; moreover, both reverse channels take $\sigma$ to $\pi$, and the inverted routing therefore implements their mixture. Consequently,
\begin{align}
&\mathcal R_{\pi\to\sigma}[(1-t)\Phi_0+t\Phi_1]\nonumber\\
&\quad=(1-t)\mathcal R_{\pi\to\sigma}[\Phi_0]+t\mathcal R_{\pi\to\sigma}[\Phi_1].
\label{eq:retrodiction-affinity}
\end{align}
The same identity holds for $t=0$ and $t=1$, so the assignment is affine on each $\Ch(\pi,\sigma)$, where both the prior and its predicted output are fixed.

The preservation of the probabilities has a simple inferential meaning: when the input is the prior, both branches predict the same output state, so observing the output provides no information about which channel was selected. The detailed circuit construction is given in the Supplemental Material.

\ 

\emph{From maximally mixed references to arbitrary priors.}
Denote the Petz assignment in Eq.~\eqref{eq:petz-main} by $\mathcal P$, and define
\begin{equation}
F_{\pi\to\sigma}[\Phi]\coloneqq\mathcal P_{\sigma\to\pi}\!\left[\mathcal R_{\pi\to\sigma}[\Phi]\right].
\label{eq:comparison-main}
\end{equation}
Composing the two reversals restores the original direction of inference. Since both assignments are involutive, $F$ is a bijection on each fixed-transition set; moreover, it preserves sequential and parallel composition, is affine, and fixes unitary channels. Uniqueness therefore amounts to showing that $F$ leaves every channel unchanged.

Maximally mixed references provide the starting point because all unitary channels on a given system preserve the same state. Every unital channel lies in their affine hull~\cite{MendlWolf2009}; affinity therefore extends the agreement from unitary channels to all unital channels. The coefficients in this affine decomposition need not be positive, but the required identity follows by rearranging it as an equality of convex mixtures. Consistency with adjoining and discarding independent systems then extends the agreement to channels connecting maximally mixed references of different dimensions.

To pass to an arbitrary full-rank reference $\rho$, we connect it to a maximally mixed state through suitable channels. Specifically, for a sufficiently large $n$, one can construct a compression channel $\mathcal C_\rho$ taking $\mathbb I_n/n$ to $\rho$ and a return channel $\mathcal D_\rho$ taking $\rho$ to $\mathbb I_n/n$, with
\begin{equation}
\mathcal C_\rho\circ\mathcal D_\rho=\frac{1}{n}\id+\frac{n-1}{n}\mathcal S_\rho\circ\Tr.
\label{eq:round-trip-main}
\end{equation}
Their explicit construction is given in the Supplemental Material. The compression admits a channel right inverse, while the return map is injective. Equation~\eqref{eq:round-trip-main} explains why affinity at arbitrary priors is useful: $F$ fixes the identity and the replacement channel $\mathcal S_\rho\circ\Tr$, hence also this composite. The opposite composite preserves a maximally mixed reference and is therefore fixed as well.

In particular, surrounding an arbitrary $\Phi\in\Ch(\pi,\sigma)$ by $\mathcal C_\pi$ and $\mathcal D_\sigma$ gives a channel between maximally mixed references, which $F$ fixes. Sequential composition and cancellation of the surrounding maps then reduce the remaining discrepancy to the form
\begin{equation}
F_{\pi\to\sigma}[\Phi]=M_\sigma\circ\Phi\circ M_\pi^{-1}.
\label{eq:reference-transformations-main}
\end{equation}
Here $M_\rho$ is an invertible channel depending only on the reference state, with $M_\rho(\rho)=\rho$. Bijectivity of $F$ and the channel right inverse of $\mathcal C_\rho$ imply that $M_\rho^{-1}$ is also a channel. Such transformations can only be unitary conjugations~\cite{BuscemiDArianoKeylPerinottiWerner2005}: $M_\rho=\Ad_{V_\rho}$, where $V_\rho$ commutes with $\rho$. Although these rotations preserve the reference states, they could still alter the predictions for other states.

The remaining question is whether these rotations can be chosen nontrivially and consistently for all reference states. Unitary covariance already restricts them to relative phases between distinct eigenspaces of the reference state, which leave the prior unchanged but can alter predictions for coherent superpositions. To constrain these phases, we adjoin independent ancillary systems whose reference states compensate for the unequal weights of different system sectors, producing joint sectors with equal reference weights. These sectors can be mixed by unitary operations without changing the joint reference, while parallel composition requires the corresponding rotation to factor into those assigned to the separate systems. Using two identical ancillas, involutivity makes their combined phase contribution independent of the sector being compared, so it cannot compensate for relative phases on the original system. Hence all these phases must agree, giving $M_\rho=\id$ and, consequently, Eq.~\eqref{eq:reference-transformations-main} gives $F[\Phi]=\Phi$, establishing uniqueness. All details are given in the Supplemental Material.


\ 

\emph{Retrodiction without involutivity.} The preceding argument also identifies the freedom that remains when involutivity is removed. Indeed, the reduction to Eq.~\eqref{eq:reference-transformations-main} relies only on (ii)--(iv), while the complete positivity of $M_\rho^{-1}$ can also be established without assuming involutivity, by a separate argument given in the Supplemental Material. This leads to the following characterization.

\begin{prop}[Characterization without involutivity]\label{lem:without-involutivity-main}
In the setting of Theorem~\ref{thm:uniqueness-main}, without assuming involutivity, an assignment satisfies (ii)--(iv) if and only if there is a family of unitary channels $M_\rho=\Ad_{V_\rho}$ such that
\begin{equation}\label{eq:phase-retrodiction-main}
\mathcal{R}_{\pi\to\sigma}[\Phi]=M_\pi\circ\mathcal{P}_{\pi\to\sigma}[\Phi]\circ M_\sigma^{-1}
\end{equation}
and
\begin{align*}
M_{U\rho U^\dagger}&=\Ad_U\circ M_\rho\circ\Ad_{U^\dagger},\\
M_{\rho\otimes\omega}&=M_\rho\otimes M_\omega
\end{align*}
for every unitary $U$ and every pair of full-rank reference states $\rho,\omega$.
\end{prop}

These assignments form a proper subclass of the doubly rotated Petz maps~\cite{SutterTomamichelHarrow2016}, since the rotations cannot be chosen independently but must satisfy the covariance and tensor-product conditions above. Reference-dependent rotations were also considered in Ref.~\cite{ParzygnatBuscemi2023}.
If we also require continuity in the prior for each fixed channel, these conditions give $M_\rho=\Ad_{\rho^{it}}$, with a single parameter $t\in\mathbb{R}$ common to all reference states, channels and systems. We then obtain the rotated Petz assignments~\cite{Wilde2015},
\begin{align}
&\mathcal R^t_{\pi\to\sigma}[\Phi](X)\nonumber\\
&\quad=\pi^{1/2+it}\,\Phi^\dagger\!\left(\sigma^{-1/2-it}X\sigma^{-1/2+it}\right)\pi^{1/2-it}.
\label{eq:rotated-petz-main}
\end{align}
Involutivity removes the remaining rotations even without assuming continuity, as shown in the proof of Theorem~\ref{thm:uniqueness-main}; within the continuous family, this selects $t=0$. The proof of the proposition and the characterization of its continuous assignments are given in the Supplemental Material.

\

\emph{The case of measurements.}
The theorem determines the action of the reverse channel on individual output states, including those associated with measurement outcomes. Let $\{E_x\}$ be a POVM, with measurement channel $\mathcal M(X)=\sum_x\Tr[E_xX]|x\rangle\langle x|$. Representing the outcome register as a quantum system keeps this example within the setting of full matrix algebras. Its reference output is diagonal, with probabilities $p_x=\Tr[\pi E_x]$. For every $p_x>0$, applying Eq.~\eqref{eq:petz-main} gives
\begin{equation}
\pi_x=\mathcal R[\mathcal M](|x\rangle\langle x|)=\frac{\sqrt{\pi}E_x\sqrt{\pi}}{p_x}.
\label{eq:measurement-retrodiction-main}
\end{equation}
Averaging these conditional states with the predicted outcome probabilities reproduces the prior, $\sum_{x:p_x>0}p_x\pi_x=\pi$, as required for reversal of the reference transition. Notice that $\pi_x$ describes inference about the input, not the physical postmeasurement state of a specified instrument. Different instruments can realize the same POVM while producing different postmeasurement states; retrodiction depends only on the measurement channel and its prior. In the case in which $\pi$ and all $E_x$ are diagonal in the same basis, Eq.~\eqref{eq:measurement-retrodiction-main} reduces to ordinary Bayesian conditioning.

Equation~\eqref{eq:measurement-retrodiction-main} is the quantum Bayes' rule for measurement outcomes obtained from the minimum-change principle in Ref.~\cite{KuangToriiBuscemi2026}, where a broad class of quantum divergences selects the same update. This rule underlies recent studies of measurement objectivity~\cite{KuangNagasawaToriiBuscemi2026}, inference beyond classical learning~\cite{Buscemi2026Learning}, and quantum $f$-divergences through retrodictive tests~\cite{Buscemi2026RetrodictiveTests}. The present theorem gives a complementary justification: any assignment satisfying (i)--(iv) for all priors and channels must use this update for measurement outcomes.


\ 

\emph{Discussion.}
The theorem justifies earlier identifications of the Petz transpose map with quantum Bayesian inversion~\cite{SuraceScandi2023,ParzygnatFullwood2023}. The choice of prior remains part of each inference problem; however, a single retrodiction assignment satisfying (i)--(iv) for all priors and channels has no further freedom. In particular, the characteristic square-root dependence on the reference states follows from operational consistency itself. This gives a compositional justification for the same prescription selected by the minimum-change principle~\cite{BaiBuscemiScarani2025}.

In statistical mechanics, Crooks arrived at the Petz transpose map for channels with an invariant reference state by extending classical Markov-chain reversal to quantum trajectories~\cite{Crooks2008}. Our theorem shows that, under the four requirements, the same prescription is uniquely determined independently of this analogy. This provides an axiomatic basis for the quantum reversal used in retrodictive formulations of fluctuation relations~\cite{KwonKim2019,BuscemiScarani2021,AwBuscemiScarani2021}.


The same conclusion applies to the retrodictive coarse-graining underlying observational entropy~\cite{BuscemiSchindlerSafranek2023,BaiEtAl2024}. Given a prior and a measurement, the outcome distribution of a state is mapped back to a quantum state by averaging the conditional states in Eq.~\eqref{eq:measurement-retrodiction-main}. Macroscopic states are precisely those reproduced by this procedure, i.e., the fixed points of measurement followed by retrodiction~\cite{NagasawaEtAl2025}. Thus, once the prior and the accessible measurement are specified, consistency fixes the retrodictive coarse-graining and hence its macroscopic states. The theorem therefore gives an independent justification for this inferential account of macroscopic description.

The Petz transpose map is also used in quantum error correction to recover encoded information after a noisy evolution~\cite{BarnumKnill2002,NgMandayam2010}, and the same recovery interpretation appears in holography. Chen, Penington, and Salton showed that the Petz transpose map suffices for entanglement-wedge reconstruction in AdS/CFT, in the semiclassical limit and for code spaces of fixed finite dimension~\cite{ChenPeningtonSalton2020}. There, information about a bulk region is recovered from the corresponding boundary region. Within a finite-dimensional channel description, our theorem fixes the retrodictive prescription, while the error-correcting properties of the holographic encoding supply the additional conditions under which that prescription also gives accurate recovery. Consistency of inference and recoverability thus explain different aspects of the same construction.

How far this characterization extends beyond full matrix algebras, in particular to general infinite-dimensional $C^*$-algebras and to the von Neumann algebraic setting~\cite{KarmakarParzygnat2026}, remains open. 
Within the present setting, however, the result explains why constructions motivated by such different physical questions converge on the same map: once retrodiction is required to satisfy the four consistency requirements for all priors and channels, the Petz transpose map is inevitable.

\medskip
\emph{Acknowledgments.}
J.S. acknowledges support from the French government under the France 2030 investment plan, as part of the Initiative d’Excellence d’Aix-Marseille Université-A*MIDEX, AMX-22-CEI-01. 
C.L. acknowledges support from the National Natural Science Foundation of China Grants No. 12575023, the Innovation Program for Quantum Science and Technology Grant No. 2021ZD0300804 and No. 2021ZD0300702. F.B. acknowledges support from MEXT Quantum Leap Flagship Program (MEXT QLEAP) Grant No.~JPMXS0120319794 and from JSPS KAKENHI Grants No.~23K03230 and~26K00621.
ChatGPT (OpenAI, Astra 6 and Sol 6.1) and Claude (Anthropic, Opus 5.5) were used, under the authors' direction, to assist with literature searches, the examination of mathematical arguments, and the preparation and revision of the manuscript. The authors take full responsibility for the content of this work.

\bibliographystyle{apsrev4-2}
\bibliography{referencesSM}

@article{BuscemiSchindlerSafranek2023,
  author        = {Buscemi, F. and Schindler, J. and {\v{S}}afr{\'a}nek, D.},
  title         = {{Observational entropy, coarse-grained states, and the Petz recovery map: information-theoretic properties and bounds}},
  journal       = {New Journal of Physics},
  volume        = {25},
  pages         = {053002},
  year          = {2023},
  doi           = {10.1088/1367-2630/accd11},
  eprint        = {2209.03803},
  archivePrefix = {arXiv},
  primaryClass  = {quant-ph},
  url           = {https://arxiv.org/abs/2209.03803}
}

@article{BaiEtAl2024,
  author        = {Bai, G. and {\v{S}}afr{\'a}nek, D. and Schindler, J. and Buscemi, F. and Scarani, V.},
  title         = {{Observational entropy with general quantum priors}},
  journal       = {Quantum},
  volume        = {8},
  pages         = {1524},
  year          = {2024},
  doi           = {10.22331/q-2024-11-14-1524},
  eprint        = {2308.08763},
  archivePrefix = {arXiv},
  primaryClass  = {quant-ph},
  url           = {https://arxiv.org/abs/2308.08763}
}

@article{NagasawaEtAl2025,
  author        = {Nagasawa, T. and Wakakuwa, E. and Kato, K. and Buscemi, F.},
  title         = {{Macroscopicity and observational deficit in states, operations, and correlations}},
  journal       = {Reports on Progress in Physics},
  volume        = {88},
  pages         = {117601},
  year          = {2025},
  doi           = {10.1088/1361-6633/ae140e},
  eprint        = {2504.12738},
  archivePrefix = {arXiv},
  primaryClass  = {quant-ph},
  url           = {https://arxiv.org/abs/2504.12738}
}

@article{Crooks2008,
  author        = {Crooks, G. E.},
  title         = {{Quantum operation time reversal}},
  journal       = {Physical Review A},
  volume        = {77},
  pages         = {034101},
  year          = {2008},
  doi           = {10.1103/PhysRevA.77.034101},
  eprint        = {0706.3749},
  archivePrefix = {arXiv},
  primaryClass  = {quant-ph},
  url           = {https://arxiv.org/abs/0706.3749}
}

@article{ChenPeningtonSalton2020,
  author        = {Chen, C.-F. and Penington, G. and Salton, G.},
  title         = {{Entanglement Wedge Reconstruction using the Petz Map}},
  journal       = {Journal of High Energy Physics},
  volume        = {2020},
  number        = {1},
  pages         = {168},
  year          = {2020},
  doi           = {10.1007/JHEP01(2020)168},
  eprint        = {1902.02844},
  archivePrefix = {arXiv},
  primaryClass  = {hep-th},
  url           = {https://arxiv.org/abs/1902.02844}
}

@misc{KuangNagasawaToriiBuscemi2026,
  author        = {Kuang, J. and Nagasawa, T. and Torii, K. and Buscemi, F.},
  title         = {{Quantifying Measurement Objectivity: A Retrodictive Approach}},
  year          = {2026},
  eprint        = {2608.17410},
  archivePrefix = {arXiv},
  primaryClass  = {quant-ph},
  url           = {https://arxiv.org/abs/2608.17410}
}

@misc{Buscemi2026Learning,
  author        = {Buscemi, F.},
  title         = {{Witnessing Quantum Bayesian Inference beyond Classical Learning}},
  year          = {2026},
  eprint        = {2610.09293},
  archivePrefix = {arXiv},
  primaryClass  = {quant-ph},
  url           = {https://arxiv.org/abs/2610.09293}
}

@misc{Buscemi2026RetrodictiveTests,
  author        = {Buscemi, F.},
  title         = {{Quantum $f$-divergences from retrodictive tests: a unifying classical variational principle}},
  year          = {2026},
  eprint        = {2610.29696},
  archivePrefix = {arXiv},
  primaryClass  = {quant-ph},
  url           = {https://arxiv.org/abs/2610.29696}
}

@article{BuscemiScarani2021,
  author        = {Buscemi, F. and Scarani, V.},
  title         = {{Fluctuation theorems from Bayesian retrodiction}},
  journal       = {Physical Review E},
  volume        = {103},
  pages         = {052111},
  year          = {2021},
  doi           = {10.1103/PhysRevE.103.052111},
  eprint        = {2009.02849},
  archivePrefix = {arXiv},
  primaryClass  = {quant-ph},
  url           = {https://arxiv.org/abs/2009.02849}
}

@misc{Parzygnat2024,
  author        = {Parzygnat, A. J.},
  title         = {{Reversing information flow: retrodiction in semicartesian categories}},
  year          = {2024},
  eprint        = {2401.17447},
  archivePrefix = {arXiv},
  primaryClass  = {quant-ph},
  url           = {https://arxiv.org/abs/2401.17447}
}

@article{KuangToriiBuscemi2026,
  author        = {Kuang, J. and Torii, K. and Buscemi, F.},
  title         = {{Quantum measurement retrodiction and entropic uncertainty relations}},
  journal       = {Quantum Science and Technology},
  volume        = {11},
  pages         = {025049},
  year          = {2026},
  doi           = {10.1088/2058-9565/ae5fcd},
  eprint        = {2511.20281},
  archivePrefix = {arXiv},
  url           = {https://arxiv.org/abs/2511.20281}
}

@book{Jaynes2003,
	address = {Cambridge},
	author = {Jaynes, E. T.},
	doi = {10.1017/CBO9780511790423},
	note = {Edited by G. L. Bretthorst},
	publisher = {Cambridge University Press},
	title = {{Probability Theory: The Logic of Science}},
	url = {https://www.cambridge.org/core/books/probability-theory/9CA08E224FF30123304E6D8935CF1A99},
	year = {2003}}

@article{DeFinetti1937,
	author = {{de Finetti}, B.},
	journal = {Annales de l'Institut Henri Poincar{\'e}},
	number = {1},
	pages = {1--68},
	title = {{La pr{\'e}vision : ses lois logiques, ses sources subjectives}},
	url = {https://www.numdam.org/item/AIHP_1937__7_1_1_0/},
	volume = {7},
	year = {1937}}

@article{BuscemiDArianoKeylPerinottiWerner2005,
  author        = {Buscemi, F. and D'Ariano, G. M. and Keyl, M. and Perinotti, P. and Werner, R. F.},
  title         = {{Clean positive operator valued measures}},
  journal       = {Journal of Mathematical Physics},
  volume        = {46},
  pages         = {082109},
  year          = {2005},
  doi           = {10.1063/1.2008996},
  eprint        = {quant-ph/0505095},
  archivePrefix = {arXiv},
  url           = {https://arxiv.org/abs/quant-ph/0505095}
}

@article{AwBuscemiScarani2021,
  author        = {Aw, C. C. and Buscemi, F. and Scarani, V.},
  title         = {{Fluctuation theorems with retrodiction rather than reverse processes}},
  journal       = {AVS Quantum Science},
  volume        = {3},
  pages         = {045601},
  year          = {2021},
  doi           = {10.1116/5.0060893},
  eprint        = {2106.08589},
  archivePrefix = {arXiv},
  url           = {https://arxiv.org/abs/2106.08589}
}

@article{ParzygnatBuscemi2023,
  author        = {Parzygnat, A. J. and Buscemi, F.},
  title         = {{Axioms for retrodiction: achieving time-reversal symmetry with a prior}},
  journal       = {Quantum},
  volume        = {7},
  pages         = {1013},
  year          = {2023},
  doi           = {10.22331/q-2023-05-23-1013},
  eprint        = {2210.13531},
  archivePrefix = {arXiv},
  url           = {https://arxiv.org/abs/2210.13531}
}

@article{MendlWolf2009,
  author        = {Mendl, C. B. and Wolf, M. M.},
  title         = {{Unital Quantum Channels: Convex Structure and Revivals of Birkhoff's Theorem}},
  journal       = {Communications in Mathematical Physics},
  volume        = {289},
  pages         = {1057--1086},
  year          = {2009},
  doi           = {10.1007/s00220-009-0824-2},
  eprint        = {0806.2820},
  archivePrefix = {arXiv},
  url           = {https://arxiv.org/abs/0806.2820}
}

@article{Leifer2006,
  author        = {Leifer, M. S.},
  title         = {{Quantum dynamics as an analog of conditional probability}},
  journal       = {Physical Review A},
  volume        = {74},
  pages         = {042310},
  year          = {2006},
  doi           = {10.1103/PhysRevA.74.042310},
  eprint        = {quant-ph/0606022v2},
  archivePrefix = {arXiv},
  url           = {https://arxiv.org/abs/quant-ph/0606022v2}
}

@article{LeiferSpekkens2013,
  author        = {Leifer, M. S. and Spekkens, R. W.},
  title         = {{Towards a formulation of quantum theory as a causally neutral theory of Bayesian inference}},
  journal       = {Physical Review A},
  volume        = {88},
  pages         = {052130},
  year          = {2013},
  doi           = {10.1103/PhysRevA.88.052130},
  eprint        = {1107.5849v4},
  archivePrefix = {arXiv},
  url           = {https://arxiv.org/abs/1107.5849v4}
}

@article{SuraceScandi2023,
  author        = {Surace, J. and Scandi, M.},
  title         = {{State retrieval beyond Bayes' retrodiction}},
  journal       = {Quantum},
  volume        = {7},
  pages         = {990},
  year          = {2023},
  doi           = {10.22331/q-2023-04-27-990},
  eprint        = {2201.09899v3},
  archivePrefix = {arXiv},
  url           = {https://arxiv.org/abs/2201.09899v3}
}

@article{ParzygnatFullwood2023,
  author        = {Parzygnat, A. J. and Fullwood, J.},
  title         = {{From time-reversal symmetry to quantum Bayes' rules}},
  journal       = {PRX Quantum},
  volume        = {4},
  pages         = {020334},
  year          = {2023},
  doi           = {10.1103/PRXQuantum.4.020334},
  eprint        = {2212.08088},
  archivePrefix = {arXiv},
  url           = {https://arxiv.org/abs/2212.08088}
}

@article{Cox1946,
  author  = {Cox, R. T.},
  title   = {{Probability, Frequency and Reasonable Expectation}},
  journal = {American Journal of Physics},
  volume  = {14},
  number  = {1},
  pages   = {1--13},
  year    = {1946},
  doi     = {10.1119/1.1990764}
}

@article{Williams1980,
  author  = {Williams, P. M.},
  title   = {{Bayesian Conditionalisation and the Principle of Minimum Information}},
  journal = {The British Journal for the Philosophy of Science},
  volume  = {31},
  number  = {2},
  pages   = {131--144},
  year    = {1980},
  doi     = {10.1093/bjps/31.2.131}
}

@article{DiaconisZabell1982,
  author  = {Diaconis, P. and Zabell, S. L.},
  title   = {{Updating Subjective Probability}},
  journal = {Journal of the American Statistical Association},
  volume  = {77},
  number  = {380},
  pages   = {822--830},
  year    = {1982},
  doi     = {10.1080/01621459.1982.10477893}
}

@article{BaiBuscemiScarani2025,
  author        = {Bai, G. and Buscemi, F. and Scarani, V.},
  title         = {{Quantum Bayes' rule and Petz transpose map from the minimum change principle}},
  journal       = {Physical Review Letters},
  volume        = {135},
  pages         = {090203},
  year          = {2025},
  doi           = {10.1103/5n4p-bxhm},
  eprint        = {2410.00319},
  archivePrefix = {arXiv},
  url           = {https://arxiv.org/abs/2410.00319}
}

@misc{KarmakarParzygnat2026,
  author        = {Karmakar, P. and Parzygnat, A. J.},
  title         = {{Bayesian inference and retrodiction for faithful states on von Neumann algebras}},
  year          = {2026},
  eprint        = {2608.20001},
  archivePrefix = {arXiv},
  primaryClass  = {math.OA},
  url           = {https://arxiv.org/abs/2608.20001}
}

@article{Petz1986,
  author  = {Petz, D.},
  title   = {{Sufficient subalgebras and the relative entropy of states of a von Neumann algebra}},
  journal = {Communications in Mathematical Physics},
  volume  = {105},
  pages   = {123--131},
  year    = {1986},
  doi     = {10.1007/BF01212345}
}

@article{Petz1988,
  author  = {Petz, D.},
  title   = {{Sufficiency of channels over von Neumann algebras}},
  journal = {The Quarterly Journal of Mathematics},
  volume  = {39},
  number  = {1},
  pages   = {97--108},
  year    = {1988},
  doi     = {10.1093/qmath/39.1.97}
}

@article{HorsmanEtAl2017,
  author        = {Horsman, D. and Heunen, C. and Pusey, M. F. and Barrett, J. and Spekkens, R. W.},
  title         = {{Can a quantum state over time resemble a quantum state at a single time?}},
  journal       = {Proceedings of the Royal Society A},
  volume        = {473},
  pages         = {20170395},
  year          = {2017},
  doi           = {10.1098/rspa.2017.0395},
  eprint        = {1607.03637},
  archivePrefix = {arXiv},
  url           = {https://arxiv.org/abs/1607.03637}
}

@article{FullwoodParzygnat2022,
  author        = {Fullwood, J. and Parzygnat, A. J.},
  title         = {{On quantum states over time}},
  journal       = {Proceedings of the Royal Society A},
  volume        = {478},
  pages         = {20220104},
  year          = {2022},
  doi           = {10.1098/rspa.2022.0104},
  eprint        = {2202.03607},
  archivePrefix = {arXiv},
  url           = {https://arxiv.org/abs/2202.03607}
}

@article{LieNg2024,
  author        = {Lie, S. H. and Ng, N. H. Y.},
  title         = {{Quantum state over time is unique}},
  journal       = {Physical Review Research},
  volume        = {6},
  pages         = {033144},
  year          = {2024},
  doi           = {10.1103/PhysRevResearch.6.033144},
  eprint        = {2308.12752},
  archivePrefix = {arXiv},
  url           = {https://arxiv.org/abs/2308.12752}
}

@article{ParzygnatFullwoodBuscemiChiribella2024,
  author        = {Parzygnat, A. J. and Fullwood, J. and Buscemi, F. and Chiribella, G.},
  title         = {{Virtual quantum broadcasting}},
  journal       = {Physical Review Letters},
  volume        = {132},
  pages         = {110203},
  year          = {2024},
  doi           = {10.1103/PhysRevLett.132.110203},
  eprint        = {2310.13049},
  archivePrefix = {arXiv},
  url           = {https://arxiv.org/abs/2310.13049}
}

@article{Wilde2015,
  author = {Wilde, Mark M.},
  title = {Recoverability in quantum information theory},
  journal = {Proceedings of the Royal Society A},
  volume = {471},
  number = {2182},
  pages = {20150338},
  year = {2015},
  doi = {10.1098/rspa.2015.0338},
  eprint = {1505.04661},
  archivePrefix = {arXiv},
  primaryClass = {quant-ph}
}

@article{SutterTomamichelHarrow2016,
  author = {Sutter, David and Tomamichel, Marco and Harrow, Aram W.},
  title = {Strengthened Monotonicity of Relative Entropy via Pinched {Petz} Recovery Map},
  journal = {IEEE Transactions on Information Theory},
  volume = {62},
  number = {5},
  pages = {2907--2913},
  year = {2016},
  doi = {10.1109/TIT.2016.2545680},
  eprint = {1507.00303},
  archivePrefix = {arXiv},
  primaryClass = {quant-ph}
}

@article{BarnumKnill2002,
  author        = {Barnum, H. and Knill, E.},
  title         = {{Reversing quantum dynamics with near-optimal quantum and classical fidelity}},
  journal       = {Journal of Mathematical Physics},
  volume        = {43},
  pages         = {2097--2106},
  year          = {2002},
  doi           = {10.1063/1.1459754},
  eprint        = {quant-ph/0004088},
  archivePrefix = {arXiv},
  url           = {https://arxiv.org/abs/quant-ph/0004088v2},
  note          = {Erratum: Journal of Mathematical Physics 67, 059901 (2026), doi: 10.1063/5.0325563}
}

@article{NgMandayam2010,
  author        = {Ng, Hui Khoon and Mandayam, Prabha},
  title         = {{Simple approach to approximate quantum error correction based on the transpose channel}},
  journal       = {Physical Review A},
  volume        = {81},
  pages         = {062342},
  year          = {2010},
  doi           = {10.1103/PhysRevA.81.062342},
  eprint        = {0909.0931},
  archivePrefix = {arXiv},
  url           = {https://arxiv.org/abs/0909.0931}
}

@article{KwonKim2019,
  author        = {Kwon, Hyukjoon and Kim, M. S.},
  title         = {{Fluctuation Theorems for a Quantum Channel}},
  journal       = {Physical Review X},
  volume        = {9},
  pages         = {031029},
  year          = {2019},
  doi           = {10.1103/PhysRevX.9.031029},
  eprint        = {1810.03150},
  archivePrefix = {arXiv},
  url           = {https://arxiv.org/abs/1810.03150}
}

@book{Rudin1987,
  author    = {Rudin, Walter},
  title     = {Real and Complex Analysis},
  edition   = {3},
  publisher = {McGraw-Hill},
  year      = {1987}
}

\clearpage
\onecolumngrid

\makeatletter
\onecolumn@grid@setup
\let\set@footnotewidth\set@footnotewidth@one
\let\compose@footnotes\compose@footnotes@one
\do@output@MVL{\global\count\footins=1000\relax}
\makeatother

\interfootnotelinepenalty=10000\relax

\appendix*

\setcounter{theorem}{0}
\setcounter{prop}{0}
\setcounter{page}{1}
\renewcommand{\thepage}{Inevitability of the Petz Transpose Map: Supplemental Material -- \arabic{page}/9}

\renewcommand{\theHtheorem}{appendix.theorem.\arabic{theorem}} \renewcommand{\theHprop}{appendix.proposition.\arabic{prop}}

\section*{Uniqueness of Petz retrodiction: Complete Proof}

For full-rank reference states $\pi$ on $\mathcal{H}_A$ and $\sigma$ on $\mathcal{H}_B$, we denote the set of channels with this reference transition by
\begin{equation*}
\Ch(\pi,\sigma)\coloneqq\bigl\{\Phi:\mathcal{B}(\mathcal{H}_A)\to\mathcal{B}(\mathcal{H}_B)\,\big|\,\Phi\text{ is a channel and }\Phi(\pi)=\sigma\bigr\}.
\end{equation*}
We also use the map associated with a full-rank state $\rho$,
\begin{equation*}
\mathbb{J}_\rho(X)\coloneqq\sqrt{\rho}X\sqrt{\rho}.
\end{equation*}
Its inverse is $\mathbb{J}_\rho^{-1}(X)=\rho^{-1/2}X\rho^{-1/2}$.

\begin{theorem}[Uniqueness of Petz retrodiction]\label{thm:uniqueness}
Suppose that, for every pair of full-rank states on arbitrary finite-dimensional systems, an assignment
\begin{equation*}
\mathcal{R}_{\pi\to\sigma}:\Ch(\pi,\sigma)\longrightarrow\Ch(\sigma,\pi)
\end{equation*}
is given. Assume that this family satisfies the following conditions.
\begin{enumerate}
\item \emph{Involutivity.} For every $\Phi\in\Ch(\pi,\sigma)$,
\begin{equation*}
\mathcal{R}_{\sigma\to\pi}\bigl[\mathcal{R}_{\pi\to\sigma}[\Phi]\bigr]=\Phi.
\end{equation*}
\item \emph{Sequential composition.} For every $\Phi\in\Ch(\pi,\sigma)$ and $\Psi\in\Ch(\sigma,\eta)$,
\begin{equation*}
\mathcal{R}_{\pi\to\eta}[\Psi\circ\Phi]=\mathcal{R}_{\pi\to\sigma}[\Phi]\circ\mathcal{R}_{\sigma\to\eta}[\Psi].
\end{equation*}
\item \emph{Parallel composition.} For every $\Phi\in\Ch(\pi,\sigma)$ and $\Psi\in\Ch(\alpha,\beta)$,
\begin{equation*}
\mathcal{R}_{\pi\otimes\alpha\to\sigma\otimes\beta}[\Phi\otimes\Psi]=\mathcal{R}_{\pi\to\sigma}[\Phi]\otimes\mathcal{R}_{\alpha\to\beta}[\Psi].
\end{equation*}
\item \emph{Unitary inversion.} For every unitary $U:\mathcal{H}_A\to\mathcal{H}_B$ and full-rank $\pi$, writing $\Ad_U(X)=UXU^\dagger$, we have
\begin{equation*}
\mathcal{R}_{\pi\to U\pi U^\dagger}[\Ad_U]=\Ad_{U^\dagger}.
\end{equation*}
\end{enumerate}
Then, for every full-rank pair $\pi,\sigma$ and every $\Phi\in\Ch(\pi,\sigma)$, with $\Phi^\dagger$ denoting the Hilbert--Schmidt adjoint, the assignment must coincide with the Petz assignment, that is,
\begin{equation}\label{eq:petz}
\mathcal{R}_{\pi\to\sigma}[\Phi]=\mathbb{J}_\pi\circ\Phi^\dagger\circ\mathbb{J}_\sigma^{-1}\eqcolon\mathcal{P}_{\pi\to\sigma}[\Phi].
\end{equation}
\end{theorem}

\

\noindent\emph{Remark (restriction to the reference supports).} In classical Bayesian inference, the prior assigns probabilities to the possible inputs of a process, and Bayes' rule updates these probabilities after we observe an output. Any input assigned zero probability by the prior keeps zero probability after the update, so inference stays within the inputs to which the prior assigns nonzero probability. Bayes' rule specifies this update only when the observed output has nonzero probability according to the prior and the process. The corresponding quantum convention is to regard the prior $\pi$ as full-rank on its support. Notice that this does not require the predicted state $\sigma=\Phi(\pi)$ to be full-rank on the original output space: the channel itself permits the corresponding restriction of that space.

To make this point precise, let $\pi$ and $\sigma$ be arbitrary density operators on $\mathcal H_A$ and $\mathcal H_B$, respectively, let $P$ and $Q$ be their support projections, and let $\Phi:\mathcal B(\mathcal H_A)\to\mathcal B(\mathcal H_B)$ be a channel satisfying $\Phi(\pi)=\sigma$. For every positive operator $X=PXP$, there exists $c\geq0$ such that $X\leq c\pi$. Positivity of $\Phi$ therefore gives $0\leq\Phi(X)\leq c\sigma$, so $\Phi(X)$ is supported on $Q$. By linearity, the same support inclusion holds for every $X=PXP$. Consequently, $\Phi$ restricts to a channel
\begin{equation*}
\Phi_{\mathrm s}:\mathcal B(P\mathcal H_A)\longrightarrow\mathcal B(Q\mathcal H_B),\qquad \Phi_{\mathrm s}(X)=\Phi(X).
\end{equation*}
Complete positivity and trace preservation are inherited from $\Phi$, while $\pi$ and $\sigma$ are full-rank when regarded as states on these reduced spaces. In particular, if $\pi$ is already full-rank on $\mathcal H_A$, then every output of $\Phi$ lies in $\mathcal B(Q\mathcal H_B)$. Restricting the output space therefore preserves the entire forward channel, even when $\sigma$ is not full-rank on $\mathcal H_B$.

Accordingly, when extending the theorem to arbitrary reference states, we understand $\Ch(\pi,\sigma)$ as the set of channels from $\mathcal B(P\mathcal H_A)$ to $\mathcal B(Q\mathcal H_B)$ taking $\pi$ to $\sigma$. Thus each reference pair fixes the input and output systems, and all four requirements, including involutivity, are imposed on channels between these systems. This convention agrees with the original definition whenever the reference states are full-rank. Moreover, restriction respects sequential composition, since the output of the first restricted channel lies in the reference support on which the second acts; it also respects parallel composition, since the support of a tensor product of states is the tensor product of their supports. Finally, a unitary mapping one reference state to another restricts to a unitary between their supports.

Theorem~\ref{thm:uniqueness} therefore applies to these reduced systems without modification, giving
\begin{equation*}
\mathcal R_{\pi\to\sigma}[\Phi_{\mathrm s}](Y)=\sqrt{\pi}\,\Phi_{\mathrm s}^{\dagger}\!\left(\sigma^{-1/2}Y\sigma^{-1/2}\right)\sqrt{\pi},\qquad Y\in\mathcal B(Q\mathcal H_B),
\end{equation*}
where $\Phi_{\mathrm s}^{\dagger}$ is the Hilbert--Schmidt adjoint between the reduced operator spaces and $\sigma^{-1/2}$ is computed on $Q\mathcal H_B$. In this precise sense, the full-rank formulation covers arbitrary priors and their predicted outputs within the present setting, with channels understood as maps between the reference supports. Equality of the restrictions of two channels to the input reference support is precisely the notion of dilational equality used in Ref.~\cite[Proposition~3.15 and Theorem~4.12]{Parzygnat2024} to formulate Petz retrodiction for nonfaithful states.

\subsection*{Proof of Theorem~\ref{thm:uniqueness}}

The Petz assignment $\mathcal{P}$ in \eqref{eq:petz} maps $\Ch(\pi,\sigma)$ to $\Ch(\sigma,\pi)$ and satisfies the four conditions, as shown in Ref.~\cite[Theorem~3.3, Proposition~5.2, and Lemma~6.1]{ParzygnatBuscemi2023}. For uniqueness, we divide the proof into steps: we first derive affinity by realizing a random choice of channels as a circuit whose reversal is fixed by the composition and unitary-inversion requirements. We then compare $\mathcal R$ with $\mathcal P$ and establish that they agree at maximally mixed priors, where all unitary channels share the same reference transition. To pass to arbitrary priors, we use a compression channel and a return channel whose composite is a mixture of the identity and preparation of the prior. Affinity at that prior fixes this composite and allows us to reduce any remaining difference to unitary conjugations depending on the reference states, which involutivity and parallel composition finally force to be the identity.

\subsubsection*{Step 1. Derive affinity from circuit compatibility.}

We begin with two elementary operations, namely preparing a reference state and discarding a system. Their retrodictions are already determined by the requirement that the assignment reverse the reference transition. Indeed, a one-dimensional system has only one state, denoted by $1$, so the preparation map $\mathcal S_\rho(z)=z\rho$ ($z\in\mathbb C$) is the only channel in $\Ch(1,\rho)$. Conversely, the trace is the only channel from the given system to a one-dimensional system, and hence the only element of $\Ch(\rho,1)$. Therefore,
\begin{equation*}
\mathcal R_{1\to\rho}[\mathcal S_\rho]=\Tr,\qquad \mathcal R_{\rho\to1}[\Tr]=\mathcal S_\rho.
\end{equation*}
In other words, retrodiction exchanges preparation and discarding. The same conclusion holds when these operations act only on an ancillary system. Indeed, unitary inversion fixes the identity channel on the system retained, while parallel composition allows us to reverse the two factors separately. Thus adjoining an ancilla in the state $\rho$, through $X\mapsto X\otimes\rho$, is reversed by tracing out that ancilla, and conversely. We will use these identities to reverse the circuit below.

We now show that sequential composition, parallel composition, and unitary inversion already imply affinity on every fixed-transition set. This means that, for any $\Phi_0,\Phi_1\in\Ch(\pi,\sigma)$ and $t\in[0,1]$,
\begin{equation*}
\mathcal R_{\pi\to\sigma}[(1-t)\Phi_0+t\Phi_1]=(1-t)\mathcal R_{\pi\to\sigma}[\Phi_0]+t\mathcal R_{\pi\to\sigma}[\Phi_1].
\end{equation*}
In other words, a probabilistic mixture of channels sharing the same reference transition is reversed into the corresponding mixture of reverse channels, with the original weights.

\begin{figure}[t]
\centering
\includegraphics[width=\linewidth]{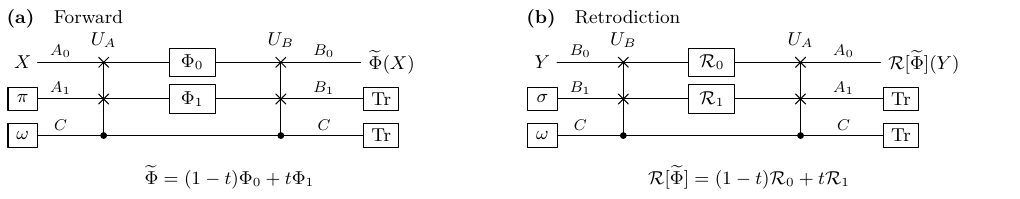}
\caption{Controlled-swap circuits for (a) the mixture $\widetilde{\Phi}$ and (b) its retrodiction. Here $\Phi_i(\pi)=\sigma$, $\mathcal{R}_i=\mathcal{R}_{\pi\to\sigma}[\Phi_i]$, and $\omega=(1-t)|0\rangle\langle0|+t|1\rangle\langle1|$, with $0<t<1$. The gates $U_A$ and $U_B$ swap the wires when the control is $|1\rangle$. Both circuits read from left to right, with arbitrary inputs $X$ and $Y$.}
\label{fig:controlled-swap-affinity}
\end{figure}

For $t=0$, both sides of this identity reduce to $\mathcal R_{\pi\to\sigma}[\Phi_0]$, while for $t=1$ they reduce to $\mathcal R_{\pi\to\sigma}[\Phi_1]$. Hence it suffices to consider $0<t<1$. We implement their mixture by routing the input through one channel while sending the common reference state through the other. Introduce two copies $A_0,A_1$ of the input system, two copies $B_0,B_1$ of the output system, and a qubit $C$ in the full-rank state $\omega=(1-t)P_0+tP_1$, where $P_i=|i\rangle\langle i|$. If $S_A$ and $S_B$ swap the corresponding copies, define the controlled swaps
\begin{equation*}
U_A=\mathbb I_{A_0A_1}\otimes P_0+S_A\otimes P_1,\qquad U_B=\mathbb I_{B_0B_1}\otimes P_0+S_B\otimes P_1.
\end{equation*}
Both are self-adjoint unitaries, and they preserve $\pi\otimes\pi\otimes\omega$ and $\sigma\otimes\sigma\otimes\omega$, respectively.  The forward and reversed circuits are shown in Fig.~\ref{fig:controlled-swap-affinity}(a) and (b), respectively.

Since $\Phi_i(\pi)=\sigma$, the mixture $\widetilde\Phi=(1-t)\Phi_0+t\Phi_1$ admits the circuit representation\footnote{For the forward channel identity, the ancillary state does not need to be $\pi$. Indeed, replacing it by any density operator $\tau$ gives, before the final partial trace, $(1-t)\Phi_0(X)\otimes\Phi_1(\tau)\otimes P_0+t\Phi_1(X)\otimes\Phi_0(\tau)\otimes P_1$. Since both channels preserve trace, the discarded factors have unit trace, and the resulting channel is still $(1-t)\Phi_0+t\Phi_1$. The choice of $\pi$ becomes important when we reverse this circuit using the axioms. With ancillary input $\tau$, the first controlled swap sends the reference state $\pi\otimes\tau\otimes\omega$ to $(1-t)\pi\otimes\tau\otimes P_0+t\tau\otimes\pi\otimes P_1$. Unless $\tau=\pi$, this state is correlated with the control qubit. Consequently, the middle channel $\Phi_0\otimes\Phi_1\otimes\id_C$ no longer acts on the product reference $\pi\otimes\pi\otimes\omega$, and parallel composition does not identify its reversal with $\mathcal R_{\pi\to\sigma}[\Phi_0]\otimes\mathcal R_{\pi\to\sigma}[\Phi_1]\otimes\id_C$. Requiring $\tau$ to be full-rank does not remove this obstruction. Thus, choosing the ancillary input to be $\pi$ ensures that the controlled swap preserves the product reference, allowing the reversal of the middle channel to be determined separately on each factor.}
\begin{equation*}
\widetilde\Phi=\Tr_{B_1C}\circ\Ad_{U_B}\circ(\Phi_0\otimes\Phi_1\otimes\id_C)\circ\Ad_{U_A}\circ(\id_{A_0}\otimes\mathcal S_\pi\otimes\mathcal S_\omega).
\end{equation*}
Indeed, in branch $i$, the input emerges as $\Phi_i(X)$ on $B_0$, while the other output is $\sigma$. All intermediate reference states are full-rank, so the composition axioms apply to every component of this circuit.

Write $\mathcal R_i=\mathcal R_{\pi\to\sigma}[\Phi_i]$. Reversing the circuit by sequential and parallel composition, and using unitary inversion together with the preparation--trace identities above, gives
\begin{equation*}
\mathcal R_{\pi\to\sigma}[\widetilde\Phi]=\Tr_{A_1C}\circ\Ad_{U_A}\circ(\mathcal R_0\otimes\mathcal R_1\otimes\id_C)\circ\Ad_{U_B}\circ(\id_{B_0}\otimes\mathcal S_\sigma\otimes\mathcal S_\omega),
\end{equation*}
where we also used the fact that the controlled swaps are self-adjoint, i.e., $U_A^\dagger=U_A$ and $U_B^\dagger=U_B$. Thus the reversed circuit is the same routing construction with the reverse channels, which satisfy $\mathcal R_i(\sigma)=\pi$. Consequently,
\begin{equation}\label{eq:derived-affinity}
\mathcal R_{\pi\to\sigma}[(1-t)\Phi_0+t\Phi_1]=(1-t)\mathcal R_{\pi\to\sigma}[\Phi_0]+t\mathcal R_{\pi\to\sigma}[\Phi_1].
\end{equation}
Thus affinity holds on every fixed-transition set, without using involutivity.

\subsubsection*{Step 2. Compare the two retrodiction assignments.}

For $\Phi\in\Ch(\pi,\sigma)$, define
\begin{equation}\label{eq:F}
F_{\pi\to\sigma}[\Phi]\coloneqq\mathcal{P}_{\sigma\to\pi}\bigl[\mathcal{R}_{\pi\to\sigma}[\Phi]\bigr].
\end{equation}
Both assignments reverse the direction of the channel and are involutive. Hence $F_{\pi\to\sigma}$ is a bijection from $\Ch(\pi,\sigma)$ to itself. Applying sequential composition twice restores the original order: for $\Phi\in\Ch(\pi,\sigma)$ and $\Psi\in\Ch(\sigma,\eta)$,
\begin{equation*}
F_{\pi\to\eta}[\Psi\circ\Phi]=F_{\sigma\to\eta}[\Psi]\circ F_{\pi\to\sigma}[\Phi].
\end{equation*}
Similarly, $F$ preserves parallel composition, is affine on every fixed-transition set, and fixes every unitary channel. It therefore suffices to prove $F[\Phi]=\Phi$, since involutivity of $\mathcal{P}$ will then give $\mathcal{R}[\Phi]=\mathcal{P}[\Phi]$. We omit the reference subscripts whenever they are clear from the context.

\subsubsection*{Step 3. Establish the result at maximally mixed priors.}

Both $\mathcal R$ and $\mathcal P$ exchange preparation and trace, as observed in Step~1. Hence $F$ fixes them and, by parallel composition, also fixes ancillary-state preparation and partial trace.

Every unitary channel leaves the maximally mixed state unchanged. Thus all unitary channels on this system have the same reference input and output, which allows us to use affinity together with the fact that $F$ fixes each of them. Indeed, consider a unital channel $\Phi$ on a $d$-dimensional system. By Ref.~\cite[Theorem~1, item~3]{MendlWolf2009}, it admits an affine decomposition
\begin{equation*}
\Phi=\sum_k c_k\Ad_{U_k},\qquad c_k\in\mathbb{R},\qquad \sum_k c_k=1.
\end{equation*}
Affinity extends uniquely to the affine hull of the convex set $\Ch(\mathbb{I}_d/d,\mathbb{I}_d/d)$, so $F$ preserves this combination even when some coefficients are negative\footnote{This step does not require extending the domain of $F$. Indeed, writing $s=\sum_{c_k<0}|c_k|$, the decomposition of $\Phi$ can be rewritten as $\frac{1}{1+s}\Phi+\sum_{c_k<0}\frac{|c_k|}{1+s}\Ad_{U_k}=\sum_{c_k>0}\frac{c_k}{1+s}\Ad_{U_k}$. Since $\sum_k c_k=1$, both sides are convex mixtures of channels in $\Ch(\mathbb I_d/d,\mathbb I_d/d)$. Applying affinity to both mixtures and rearranging yields $F[\Phi]=\sum_k c_kF[\Ad_{U_k}]$. Thus every evaluation of $F$ remains within its original domain.}. Since each unitary channel is fixed, we obtain
\begin{equation*}
F[\Phi]=\sum_k c_kF[\Ad_{U_k}]=\sum_k c_k\Ad_{U_k}=\Phi.
\end{equation*}

To cover different input and output dimensions, let $\Phi:A\to B$ satisfy $\Phi(\mathbb{I}_A/d_A)=\mathbb{I}_B/d_B$, and define a channel on $A\otimes B$ by
\begin{equation*}
\widehat{\Phi}(Z)=\frac{\mathbb{I}_A}{d_A}\otimes\Phi(\Tr_B[Z]).
\end{equation*}
This channel fixes $\mathbb{I}_{AB}/(d_Ad_B)$, hence is unital and satisfies $F[\widehat{\Phi}]=\widehat{\Phi}$. Moreover,
\begin{equation*}
\Phi(X)=\Tr_A\!\left[\widehat{\Phi}\!\left(X\otimes\frac{\mathbb{I}_B}{d_B}\right)\right].
\end{equation*}
Thus $\Phi$ is a composition of ancillary-state preparation, $\widehat{\Phi}$, and a partial trace, all fixed by $F$. Sequential composition consequently gives
\begin{equation}\label{eq:uniform}
F[\Phi]=\Phi
\end{equation}
for every $\Phi\in\Ch(\mathbb{I}_A/d_A,\mathbb{I}_B/d_B)$.

\subsubsection*{Step 4. Reduce the difference at arbitrary priors to unitary channels.}

To pass from maximally mixed priors to an arbitrary full-rank state $\rho$ on a $d$-dimensional system, we construct a channel that performs this reference transition and has a channel right inverse. Choose an integer $n>d$ such that $n\rho\geq\mathbb I_d$, and an isometry $V:\mathbb C^d\to\mathbb C^n$. The operator
\begin{equation*}
\xi_\rho=\frac{n\rho-\mathbb I_d}{n-d}
\end{equation*}
is a density operator. Using it, define
\begin{equation*}
\mathcal C_\rho(X)=V^\dagger XV+\Tr\!\left[(\mathbb I_n-VV^\dagger)X\right]\xi_\rho.
\end{equation*}
This channel, compressing $n\times n$ matrices into $d\times d$ matrices, can be implemented by measuring the two projections $VV^\dagger$ and $\mathbb I_n-VV^\dagger$, mapping the first subspace back through $V^\dagger$, and preparing $\xi_\rho$ for the second outcome. Its construction gives
\begin{equation*}
\mathcal C_\rho(\mathbb I_n/n)=\rho,\qquad
\mathcal C_\rho\circ\Ad_V=\id,
\end{equation*}
where $\Ad_V(X)=VXV^\dagger$. In particular, $\mathcal C_\rho\in\Ch(\mathbb I_n/n,\rho)$.

For the opposite reference transition, define the channel
\begin{equation*}
\mathcal D_\rho(X)=\frac{1}{n}\left(VXV^\dagger+\Tr[X](\mathbb I_n-V\rho V^\dagger)\right).
\end{equation*}
Indeed, $\mathbb I_n-V\rho V^\dagger\geq0$ has trace $n-1$, so $\mathcal D_\rho$ is a mixture of the isometric embedding and preparation of $(\mathbb I_n-V\rho V^\dagger)/(n-1)$. It satisfies $\mathcal D_\rho(\rho)=\mathbb I_n/n$, and its composite with the compression channel is
\begin{equation*}
\mathcal C_\rho\circ\mathcal D_\rho
=\frac{1}{n}\id+\frac{n-1}{n}\mathcal S_\rho\circ\Tr.
\end{equation*}
Both channels in this mixture preserve $\rho$ and are fixed by $F$. Hence affinity at $\rho$ fixes $\mathcal C_\rho\circ\mathcal D_\rho$. The opposite composite $\mathcal D_\rho\circ\mathcal C_\rho$ preserves the maximally mixed state and is fixed by \eqref{eq:uniform}. Sequential composition therefore gives
\begin{equation*}
F[\mathcal C_\rho]\circ F[\mathcal D_\rho]=\mathcal C_\rho\circ\mathcal D_\rho,\qquad
F[\mathcal D_\rho]\circ F[\mathcal C_\rho]=\mathcal D_\rho\circ\mathcal C_\rho.
\end{equation*}
The first composite is invertible as a linear map: it fixes $\rho$ and acts as multiplication by $1/n$ on traceless operators. Consequently, $\mathcal D_\rho$ and $F[\mathcal D_\rho]$ are injective, and $F[\mathcal C_\rho]$ is surjective.

Define the channel
\begin{equation}\label{eq:Mrho}
M_\rho\coloneqq F[\mathcal C_\rho]\circ\Ad_V.
\end{equation}
Composing the second identity above on the right with $\Ad_V$ gives $F[\mathcal D_\rho]\circ M_\rho=\mathcal D_\rho$. Consequently,
\begin{equation*}
F[\mathcal D_\rho]\circ M_\rho\circ\mathcal C_\rho
=\mathcal D_\rho\circ\mathcal C_\rho
=F[\mathcal D_\rho]\circ F[\mathcal C_\rho],
\end{equation*}
and injectivity of $F[\mathcal D_\rho]$ gives $F[\mathcal C_\rho]=M_\rho\circ\mathcal C_\rho$. Since $F[\mathcal C_\rho]$ is surjective, the square map $M_\rho$ is invertible. We have therefore obtained
\begin{equation*}
F[\mathcal C_\rho]=M_\rho\circ\mathcal C_\rho,\qquad
F[\mathcal D_\rho]=\mathcal D_\rho\circ M_\rho^{-1}.
\end{equation*}
The first identity also gives $M_\rho(\rho)=\rho$. Moreover, $M_{\mathbb I_d/d}=\id$, since in that case \eqref{eq:uniform} fixes $\mathcal C_\rho$.

The construction above applies to every full-rank reference state $\rho$. We now use it twice, for the input reference $\pi$ and the output reference $\sigma$ of an arbitrary channel $\Phi\in\Ch(\pi,\sigma)$, to determine how $F$ acts on $\Phi$. The composite $\mathcal D_\sigma\circ\Phi\circ\mathcal C_\pi$ has maximally mixed input and output references: $\mathcal C_\pi$ first maps the maximally mixed input to $\pi$, and $\mathcal D_\sigma$ maps the output reference $\sigma$ back to the maximally mixed state. Thus \eqref{eq:uniform} and sequential composition give
\begin{align*}
\mathcal D_\sigma\circ\Phi\circ\mathcal C_\pi
&=F[\mathcal D_\sigma]\circ F[\Phi]\circ F[\mathcal C_\pi]\\
&=\mathcal D_\sigma\circ M_\sigma^{-1}\circ F[\Phi]\circ M_\pi\circ\mathcal C_\pi.
\end{align*}
Since $\mathcal D_\sigma$ is injective and $\mathcal C_\pi$ is surjective, we may cancel them on both sides and obtain
\begin{equation}\label{eq:channel-transformation}
\Phi = M_\sigma^{-1}\circ F[\Phi]\circ M_\pi\quad\iff\quad F_{\pi\to\sigma}[\Phi]=M_\sigma\circ\Phi\circ M_\pi^{-1}.
\end{equation}
In other words, the difference between the two retrodiction assignments is determined by one invertible channel for each prior.

Having determined $F$ on arbitrary channels, we return to a single reference state $\rho$ to show that $M_\rho^{-1}$ is also a channel. The compression channel $\mathcal C_\rho$ belongs to $\Ch(\mathbb I_n/n,\rho)$, and $F$ is a bijection on this set. Hence we may apply \eqref{eq:channel-transformation} to $F^{-1}[\mathcal C_\rho]$. Using $M_{\mathbb I_n/n}=\id$, we obtain $F^{-1}[\mathcal C_\rho]=M_\rho^{-1}\circ\mathcal C_\rho$. Composing on the right with $\Ad_V$ yields
\begin{equation*}
M_\rho^{-1}=F^{-1}[\mathcal C_\rho]\circ\Ad_V.
\end{equation*}
Both factors on the right are channels. Consequently, both $M_\rho$ and $M_\rho^{-1}$ are CPTP, and the classification of square channels with a CPTP inverse~\cite[Theorem III.2]{BuscemiDArianoKeylPerinottiWerner2005} gives
\begin{equation*}
M_\rho=\Ad_{V_\rho},\qquad V_\rho\text{ unitary},\qquad [V_\rho,\rho]=0.
\end{equation*}
The commutation relation follows from $M_\rho(\rho)=\rho$.

\subsubsection*{Step 5. Derive involutivity and parallel composition for $M_\rho$.}

Since $V_\rho$ commutes with $\rho$, the channel $M_\rho$ commutes with $\mathbb{J}_\rho$ and its inverse; moreover, $M_\rho^\dagger=M_\rho^{-1}$. Applying Petz retrodiction to \eqref{eq:channel-transformation} and using \eqref{eq:F} therefore gives
\begin{equation*}
\mathcal{R}_{\pi\to\sigma}[\Phi]=M_\pi\circ\mathcal{P}_{\pi\to\sigma}[\Phi]\circ M_\sigma^{-1}.
\end{equation*}
Applying the same identity to $\mathcal{R}[\Phi]$ and using involutivity of $\mathcal{P}$ yields
\begin{equation*}
\mathcal{R}_{\sigma\to\pi}\bigl[\mathcal{R}_{\pi\to\sigma}[\Phi]\bigr]
=M_\sigma^2\circ\Phi\circ M_\pi^{-2}.
\end{equation*}
By involutivity of $\mathcal R$, the preceding identity with $\Phi=\mathcal C_\rho$ reads $\mathcal C_\rho=M_\rho^2\circ\mathcal C_\rho$, since $\mathcal C_\rho$ sends $\mathbb I_n/n$ to $\rho$ and $M_{\mathbb I_n/n}=\id$. Composing on the right with its channel right inverse $\Ad_V$ gives
\begin{equation}\label{eq:involution}
M_\rho^2=\id.
\end{equation}

Next, let $\omega$ be an arbitrary full-rank state on another finite-dimensional system. Parallel composition gives $F[\mathcal C_\rho\otimes\mathcal C_\omega]=F[\mathcal C_\rho]\otimes F[\mathcal C_\omega]$. The channel $\mathcal C_\rho\otimes\mathcal C_\omega$ has output prior $\rho\otimes\omega$, while its two factors have output priors $\rho$ and $\omega$, respectively; all three input priors are maximally mixed. Applying \eqref{eq:channel-transformation} on both sides, and recalling that $M=\id$ at maximally mixed priors, we therefore obtain
\begin{equation*}
M_{\rho\otimes\omega}\circ(\mathcal C_\rho\otimes\mathcal C_\omega)
=(M_\rho\otimes M_\omega)\circ(\mathcal C_\rho\otimes\mathcal C_\omega).
\end{equation*}
Composing on the right with the tensor product of the channel right inverses of $\mathcal C_\rho$ and $\mathcal C_\omega$ yields
\begin{equation}\label{eq:tensor}
M_{\rho\otimes\omega}=M_\rho\otimes M_\omega.
\end{equation}

\subsubsection*{Step 6. Eliminate the remaining unitary freedom.}

Let $U$ be any unitary commuting with $\rho$, and choose a unitary square root $W$ of $U$ (i.e., $W^2=U$) that also commutes with $\rho$. Since $F$ fixes unitary channels, \eqref{eq:channel-transformation} gives
\begin{equation*}
M_\rho\circ\Ad_W\circ M_\rho^{-1}=\Ad_W.
\end{equation*}
The left-hand side is $\Ad_{M_\rho(W)}$, because $M_\rho$ is a unitary conjugation. Hence $M_\rho(W)=aW$ for some scalar phase $a$. Applying $M_\rho$ again and using involutivity gives $a^2=1$. Since $M_\rho$ preserves multiplication, we therefore obtain
\begin{equation*}
M_\rho(U)=M_\rho(W^2)=M_\rho(W)^2=a^2W^2=U.
\end{equation*}
Thus $V_\rho$ (i.e., the unitary corresponding to the channel $M_\rho$) commutes with every unitary commuting with $\rho$, and hence is scalar on each spectral subspace of $\rho$. Writing the distinct eigenvalues of $\rho$ as $r_j$, we have
\begin{equation*}
\rho=\sum_j r_jP_j,\qquad
V_\rho=\sum_j v_jP_j,\qquad |v_j|=1,
\end{equation*}
where $P_j$ projects onto the whole eigenspace associated with $r_j$.

It remains to show that these scalar phases agree. For this purpose, we introduce \textit{two} ancillary copies of the system, each with the full-rank reference state
\begin{equation*}
\omega=\frac{\rho^{-1/2}}{c},
\qquad c=\Tr[\rho^{-1/2}].
\end{equation*}
Thus the joint reference state is $\rho\otimes\omega\otimes\omega$. The inverse square root is chosen so that all the subspaces with projections $P_j\otimes P_j\otimes P_j$ belong to the same eigenspace of this state: their common eigenvalue is
\begin{equation*}
r_j\left(\frac{r_j^{-1/2}}{c}\right)^2=\frac{1}{c^2}.
\end{equation*}
The state $\omega$ has the same spectral projections as $\rho$, so the preceding argument gives $V_\omega=\sum_j t_jP_j$. Moreover, involutivity~\eqref{eq:involution} implies that $V_\omega^2=z\mathbb I$ for some scalar phase $z$, and therefore $t_j^2=z$ for every $j$.

Applying \eqref{eq:tensor} twice, once for each ancillary copy, gives
\begin{equation*}
M_{\rho\otimes\omega\otimes\omega}
=M_{\rho\otimes\omega}\otimes M_\omega
=M_\rho\otimes M_\omega\otimes M_\omega.
\end{equation*}
Thus the unitary $V_\rho\otimes V_\omega\otimes V_\omega$ implements $M_{\rho\otimes\omega\otimes\omega}$, and the preceding commutant argument implies that it is scalar on the entire common eigenspace identified above. On the subspace with projection $P_j\otimes P_j\otimes P_j$, however, its scalar value is $v_jt_j^2=zv_j$. Denoting its scalar value on this common eigenspace by $w$, we therefore have $zv_j=w$ for every $j$. Since $z\neq0$, it follows that $v_j$ is independent of $j$.

Consequently, $V_\rho$ is scalar and $M_\rho=\id$ for every full-rank prior. Equation~\eqref{eq:channel-transformation} gives $F[\Phi]=\Phi$, and \eqref{eq:F} then gives $\mathcal R[\Phi]=\mathcal P[\Phi]$.

\section*{Characterization without involutivity: Complete Proof}

\begin{prop}[Characterization without involutivity]\label{lem:without-involutivity}
In the setting of Theorem~\ref{thm:uniqueness}, without assuming involutivity, an assignment satisfies sequential composition, parallel composition, and unitary inversion if and only if there is a family of unitary channels $M_\rho=\Ad_{V_\rho}$ such that
\begin{equation}\label{eq:phase-retrodiction}
\mathcal R_{\pi\to\sigma}[\Phi]=M_\pi\circ\mathcal P_{\pi\to\sigma}[\Phi]\circ M_\sigma^{-1}
\end{equation}
and:
\begin{align}
M_{U\rho U^\dagger}&=\Ad_U\circ M_\rho\circ\Ad_{U^\dagger},\label{eq:reference-unitary-covariance}\\
M_{\rho\otimes\omega}&=M_\rho\otimes M_\omega,\label{eq:reference-unitary-tensor}
\end{align}
for every unitary $U$ and every pair of full-rank states $\rho,\omega$. The family of channels $M_\rho$ is uniquely determined by $\mathcal R$, whereas each implementing unitary $V_\rho$ is determined only up to an overall phase. Equation~\eqref{eq:phase-retrodiction} has the doubly rotated Petz form of Ref.~\cite{SutterTomamichelHarrow2016}, with the rotations constrained by the compatibility conditions above.

Among these assignments, those continuous in the prior for each fixed channel are precisely the rotated Petz assignments~\cite{Wilde2015}
\begin{equation}\label{eq:rotated-petz}
\mathcal R^t_{\pi\to\sigma}[\Phi](X)=\pi^{1/2+it}\,\Phi^\dagger\!\left(\sigma^{-1/2-it}X\sigma^{-1/2+it}\right)\pi^{1/2-it},\qquad t\in\mathbb{R}.
\end{equation}
Here continuity means that, for every fixed channel $\Phi$, the map $\pi\mapsto\mathcal R_{\pi\to\Phi(\pi)}[\Phi]$ is continuous on the set of states for which both $\pi$ and $\Phi(\pi)$ are full-rank. The parameter $t$ is unique and is the same for all reference states, channels and systems.
\end{prop}

\subsection*{Proof of Proposition~\ref{lem:without-involutivity}}

Steps~1--4 of the proof of Theorem~\ref{thm:uniqueness}, up to \eqref{eq:channel-transformation}, remain valid after omitting the assertion that $F$ is bijective. Thus, with $F$ defined in \eqref{eq:F}, we have
\begin{equation}\label{eq:noinv-channel-transformation}
F_{\pi\to\sigma}[\Phi]=M_\sigma\circ\Phi\circ M_\pi^{-1},
\end{equation}
where each $M_\rho$ is invertible as a linear channel satisfying $M_\rho(\rho)=\rho$, and $M_{\mathbb{I}_d/d}=\id$.

It remains to show that $M_\rho^{-1}$ is completely positive without using $F^{-1}$. Let $E_\rho=\Ad_{W_\rho}$ be the explicit channel right inverse of $\mathcal C_\rho$ constructed in Step~4, where $W_\rho$ is the isometry denoted there by $V$. Thus $\mathcal C_\rho\circ E_\rho=\id$. Applying \eqref{eq:noinv-channel-transformation} to $\Phi=\mathcal P[\mathcal C_\rho]$, and using the Petz formula and the involutivity of $\mathcal P$, gives
\begin{equation*}
K_\rho:=\mathcal R[\mathcal P[\mathcal C_\rho]]\circ E_\rho=\mathbb{J}_\rho\circ(M_\rho^{-1})^\dagger\circ\mathbb{J}_\rho^{-1}.
\end{equation*}
Since $K_\rho$ is a channel,
\begin{equation*}
M_\rho^{-1}=\mathbb{J}_\rho\circ K_\rho^\dagger\circ\mathbb{J}_\rho^{-1}
\end{equation*}
is completely positive. Its trace preservation follows from that of $M_\rho$. The classification of channels with a CPTP inverse used in Step~4 therefore gives $M_\rho=\Ad_{V_\rho}$ with $[V_\rho,\rho]=0$.

Applying Petz retrodiction to \eqref{eq:noinv-channel-transformation}, as at the beginning of Step~5, yields \eqref{eq:phase-retrodiction}. Since $F$ fixes unitary channels, the same transformation formula gives \eqref{eq:reference-unitary-covariance}. The tensor-composition argument of Step~5 also applies unchanged and gives \eqref{eq:reference-unitary-tensor}. This proves the forward implication in the first part of the proposition.

Before proving the converse, we show that, for any family of unitary channels, the covariance condition $M_{U\rho U^\dagger}=\Ad_U\circ M_\rho\circ\Ad_{U^\dagger}$ alone implies that every operator commuting with $\rho$ is a fixed point of $M_\rho$. This consequence of covariance will also be used in the uniqueness and continuity arguments below. Let $U$ be a unitary commuting with $\rho$, and write its spectral decomposition as
\begin{equation*}
U=\sum_j e^{i\theta_j}Q_j,\qquad \theta_j\in(-\pi,\pi].
\end{equation*}
Define the unitary cube root
\begin{equation*}
W:=\sum_j e^{i\theta_j/3}Q_j.
\end{equation*}
As a spectral function of $U$, the operator $W$ also commutes with $\rho$, and $W^3=U$. This choice ensures that its trace is nonzero. Indeed, if the system has dimension $d$, then
\begin{equation*}
\operatorname{Re}\Tr W=\sum_j\cos(\theta_j/3)\Tr Q_j\geq\frac12\sum_j\Tr Q_j=\frac d2>0,
\end{equation*}
where in the first inequality we used the fact that for every $j$ $\theta_j\in(-\pi,\pi]$.
Since $W\rho W^\dagger=\rho$, applying the covariance condition~\eqref{eq:reference-unitary-covariance} with $U=W$ gives $M_\rho=\Ad_W\circ M_\rho\circ\Ad_{W^\dagger}$, or equivalently $M_\rho\circ\Ad_W\circ M_\rho^{-1}=\Ad_W$. Since $M_\rho$ is a unitary conjugation, the left-hand side is $\Ad_{M_\rho(W)}$. Hence $M_\rho(W)=aW$ for a scalar $a$ with $|a|=1$. Taking traces gives $\Tr W=a\Tr W$, and the nonzero trace just established implies $a=1$. Unitary conjugation preserves products, so
\begin{equation*}
M_\rho(U)=M_\rho(W^3)=M_\rho(W)^3=W^3=U.
\end{equation*}
It follows that $V_\rho$ commutes with every unitary commuting with $\rho$. Consequently, for the spectral decomposition of $\rho$ into distinct eigenvalues,
\begin{equation*}
\rho=\sum_j r_jP_j,\qquad V_\rho=\sum_j v_jP_j,\qquad |v_j|=1,
\end{equation*}
where each $P_j$ projects onto the whole corresponding eigenspace. Thus $M_\rho(X)=X$ whenever $[X,\rho]=0$. In particular, $[V_\rho,\rho]=0$ and $M_{\mathbb{I}_d/d}=\id$.

We now prove the converse. Take any family of unitary channels satisfying \eqref{eq:reference-unitary-covariance} and \eqref{eq:reference-unitary-tensor}, and define $\mathcal R$ by \eqref{eq:phase-retrodiction}. The preceding argument shows that each $M_\rho$ fixes $\rho$. Hence $\mathcal R_{\pi\to\sigma}[\Phi]$ is a channel mapping $\sigma$ to $\pi$. Sequential composition follows by cancellation of the intermediate terms $M_\sigma^{-1}\circ M_\sigma$. Parallel composition follows from \eqref{eq:reference-unitary-tensor}. For $\sigma=U\pi U^\dagger$, the identity $M_\sigma=\Ad_U\circ M_\pi\circ\Ad_{U^\dagger}$ and the unitary-inversion property of $\mathcal P$ give
\begin{equation*}
M_\pi\circ\Ad_{U^\dagger}\circ M_\sigma^{-1}=\Ad_{U^\dagger},
\end{equation*}
which proves unitary inversion for $\mathcal R$.

We next prove that the representing family of channels $M_\rho$ is unique. For any family representing $\mathcal R$, applying Petz retrodiction to \eqref{eq:phase-retrodiction} recovers \eqref{eq:noinv-channel-transformation}. Write $u_n:=\mathbb{I}_n/n$, where $n$ is the input dimension of the compression channel $\mathcal C_\rho$ constructed in Step~4 of the proof of Theorem~\ref{thm:uniqueness}. The equality $M_{u_n}=\id$ removes the input-side factor from the transformation formula, giving
$F_{u_n\to\rho}[\mathcal C_\rho]=M_\rho\circ\mathcal C_\rho$.
Composing with $E_\rho$ and using $\mathcal C_\rho\circ E_\rho=\id$ therefore gives
\begin{equation*}
M_\rho=F_{u_n\to\rho}[\mathcal C_\rho]\circ E_\rho.
\end{equation*}
This determines $M_\rho$ uniquely from $\mathcal R$. Indeed, $F$ is determined by $\mathcal R$, and the same auxiliary channels $\mathcal C_\rho$ and $E_\rho$ can be used for any representing family, so every such family gives the same $M_\rho$. The implementing unitary remains free up to an overall phase. This proves the uniqueness of the representing family of channels and completes the proof of the first part of the proposition.

\medskip
We now show that the assignments continuous in the prior are the rotated Petz assignments. For $s\in\mathbb{R}$, define the qubit state and its associated channel
\begin{equation*}
\tau_s:=\frac{\operatorname{diag}(e^s,1)}{1+e^s},\qquad \mathcal Q_s:=M_{\tau_s}.
\end{equation*}
As proved above, the unitary implementing $\mathcal Q_s$ acts by a phase on each eigenspace of $\tau_s$. Since $\tau_s$ is diagonal in the computational basis, so is this unitary. Therefore,
\begin{equation*}
\mathcal Q_s(|0\rangle\langle1|)=q(s)|0\rangle\langle1|,\qquad |q(s)|=1,
\end{equation*}
and $q(0)=1$ because $\tau_0=\mathbb{I}_2/2$. Tensor compatibility relates these phases for different priors. The basis vectors $|001\rangle$ and $|110\rangle$ are eigenvectors of the joint density operator $\Omega:=\tau_s\otimes\tau_u\otimes\tau_{s+u}$ with the same eigenvalue $e^{s+u}/[(1+e^s)(1+e^u)(1+e^{s+u})]$. Therefore, the operator $|001\rangle\langle110|$ commutes with $\Omega$. As proved above, $M_\Omega$ leaves every operator commuting with $\Omega$ unchanged. By tensor compatibility~\eqref{eq:reference-unitary-tensor}, $M_\Omega=\mathcal Q_s\otimes\mathcal Q_u\otimes\mathcal Q_{s+u}$. Applying this identity to $|001\rangle\langle110|$ and using $M_\Omega(|001\rangle\langle110|)=|001\rangle\langle110|$ gives $q(s)q(u)\overline{q(s+u)}=1$, and thus \begin{equation*}
q(s+u)=q(s)q(u)
\end{equation*}. 
We next derive continuity of $q$ from the stated continuity of $\mathcal R$, using one fixed channel. Let $|\pm\rangle=(|0\rangle\pm|1\rangle)/\sqrt2$ and define
\begin{equation*}
\mathcal M_X(X):=\langle+|X|+\rangle\,|0\rangle\langle0|+\langle-|X|-\rangle\,|1\rangle\langle1|.
\end{equation*}
This is a channel between full qubit matrix algebras with diagonal outputs, and $\mathcal M_X(\tau_s)=\mathbb{I}_2/2$ for every $s$. Its Petz reverse satisfies
\begin{equation*}
\mathcal P_{\tau_s\to\mathbb{I}_2/2}[\mathcal M_X](|0\rangle\langle0|)=2\sqrt{\tau_s}\,|+\rangle\langle+|\sqrt{\tau_s}.
\end{equation*}
Using \eqref{eq:phase-retrodiction}, $M_{\mathbb{I}_2/2}=\id$, and the commutation of $V_{\tau_s}$ with $\tau_s$, we obtain
\begin{equation*}
\mathcal Q_s(|+\rangle\langle+|)=\frac12\,\tau_s^{-1/2}\mathcal R_{\tau_s\to\mathbb{I}_2/2}[\mathcal M_X](|0\rangle\langle0|)\tau_s^{-1/2}.
\end{equation*}
The right-hand side is continuous in $s$. Its $(0,1)$ matrix element is $q(s)/2$, so $q$ is continuous. Since $q$ is continuous, takes values on the unit circle, and satisfies $q(s+u)=q(s)q(u)$, then $q(s)=e^{its}$ for a unique $t\in\mathbb R$~\cite{Rudin1987}.

Finally, we use this qubit family to determine $M_\rho$ for an arbitrary reference state $\rho$. Let $|a\rangle,|b\rangle$ be orthonormal eigenvectors of $\rho$ with eigenvalues $r_a,r_b>0$, and set $s=\log(r_a/r_b)$. The vectors $|a\rangle\otimes|1\rangle$ and $|b\rangle\otimes|0\rangle$ are eigenvectors of the joint density operator $\rho\otimes\tau_s$ with the same eigenvalue, $r_a/(1+e^s)=r_b e^s/(1+e^s)$. As proved above, the unitary implementing $M_{\rho\otimes\tau_s}$ acts by the same phase on these two vectors. Thus $|a\rangle\langle b|\otimes|1\rangle\langle0|$ is an eigenoperator with eigenvalue $1$ of $M_{\rho\otimes\tau_s}=M_\rho\otimes\mathcal Q_s$. Since $\mathcal Q_s(|1\rangle\langle0|)=\overline{q(s)}|1\rangle\langle0|$, we conclude that
\begin{equation*}
M_\rho(|a\rangle\langle b|)=q(s)|a\rangle\langle b|=\left(\frac{r_a}{r_b}\right)^{it}|a\rangle\langle b|.
\end{equation*}
The projectors $|a\rangle\langle a|$ are also eigenoperators of $M_\rho$, with eigenvalue $1$. Thus $M_\rho$ and $\Ad_{\rho^{it}}$ agree on the operator basis $\{|a\rangle\langle b|\}$, giving $M_\rho=\Ad_{\rho^{it}}$ in every dimension. The same parameter $t$ applies to all reference states and systems because each has been compared with the same qubit family. Substitution into \eqref{eq:phase-retrodiction} proves \eqref{eq:rotated-petz}. Conversely, for any real $t$, the identity $(U\rho U^\dagger)^{it}=U\rho^{it}U^\dagger$ shows that the family $M_\rho=\Ad_{\rho^{it}}$ satisfies unitary covariance. The identity $(\rho\otimes\omega)^{it}=\rho^{it}\otimes\omega^{it}$ gives tensor compatibility, and the resulting assignment is continuous on the full-rank domain. This concludes the proof.

\section*{Comparison with previous work}

Theorem~\ref{thm:uniqueness} establishes uniqueness under the retrodiction axioms of Ref.~\cite{ParzygnatBuscemi2023} in the setting of full matrix algebras. We find that affinity need not be imposed but can in fact be derived from sequential composition, parallel composition, and unitary inversion. The theorem therefore answers Question~7.1 of that reference in the setting of full matrix algebras.

The affinity proved above has a simple operational interpretation. Suppose a channel is selected at random from channels that all map $\pi$ to $\sigma$. When the input is $\pi$, the output is $\sigma$ regardless of which channel was selected, so measuring the output gives no information about that choice. Affinity means that the reverse channel is a mixture of the corresponding reverse channels with the same selection probabilities. Step~1 shows that this behavior is already required by compatibility with a circuit implementing the random choice, and holds on every fixed-transition set without using involutivity.


\end{document}